\documentclass[
twocolumn,
superscriptaddress,
amsmath,amssymb,
prl,
]{revtex4-2}
\usepackage{graphicx}% Include figure files
\usepackage{hyperref}
\usepackage{dcolumn}% Align table columns on decimal point
\usepackage{bm}% bold math
\usepackage{epstopdf}
\usepackage{romannum}
\usepackage{csquotes}
\usepackage[dvipsnames]{xcolor}
\usepackage{hyperref}% add hypertext capabilities
\begin{document}
	
	\preprint{FeSeS}
	
	\title{Intrinsic pinning of FeSe$_1$$_-$$_x$S$_x$ single crystals probed by torque magnetometry}% Force line breaks with \\

    \author{Nan Zhou}
    \affiliation{School of Physics, Southeast University, Nanjing 211189, China}
    \affiliation{Key Laboratory of Materials Physics, Institute of Solid State Physics, HFIPS, Chinese Academy of Sciences, Hefei 230031, China }
    \affiliation{Institute for Solid State Physics (ISSP), The University of Tokyo, Kashiwa, Chiba 277-8581, Japan}
    
   \author{Yue Sun}
   \email{Corresponding author:sunyue@seu.edu.cn}
   \affiliation{School of Physics, Southeast University, Nanjing 211189, China}

   \author{Q. Hou}
   \affiliation{School of Physics, Southeast University, Nanjing 211189, China}
   
   \author{T. Sakakibara}
   \affiliation{Institute for Solid State Physics (ISSP), The University of Tokyo, Kashiwa, Chiba 277-8581, Japan}
   
   \author{X. Z. Xing}
   \affiliation{School of Physics, Southeast University, Nanjing 211189, China}
   
   \author{C. Q. Xu}
   \affiliation{School of Physics, Southeast University, Nanjing 211189, China}
   
   \author{C. Y. Xi}
   \affiliation{Anhui Province Key Laboratory of Condensed Matter Physics at Extreme Conditions, High Magnetic Field Laboratory, HFIPS, Chinese Academy of Sciences, Hefei 230031, China}
   
   \author{Z. S. Wang}
   \affiliation{Anhui Province Key Laboratory of Condensed Matter Physics at Extreme Conditions, High Magnetic Field Laboratory, HFIPS, Chinese Academy of Sciences, Hefei 230031, China}

   \author{Y. F. Zhang}
   \affiliation{School of Physics, Southeast University, Nanjing 211189, China}
   \affiliation{Institute for Solid State Physics (ISSP), The University of Tokyo, Kashiwa, Chiba 277-8581, Japan}

   \author{Y. Q. Pan}
   \affiliation{School of Physics, Southeast University, Nanjing 211189, China}
   
   \author{B. Chen}
   \affiliation{Department of Physics and Hangzhou Key Laboratory of Quantum Matters, Hangzhou Normal University, Hangzhou 310036, China}
   
   \author{X. Luo}
   \affiliation{Key Laboratory of Materials Physics, Institute of Solid State Physics, HFIPS, Chinese Academy of Sciences, Hefei 230031, China }
   
   \author{Y. P. Sun}
   \affiliation{Key Laboratory of Materials Physics, Institute of Solid State Physics, HFIPS, Chinese Academy of Sciences, Hefei 230031, China }
   \affiliation{Anhui Province Key Laboratory of Condensed Matter Physics at Extreme Conditions, High Magnetic Field Laboratory, HFIPS, Chinese Academy of Sciences, Hefei 230031, China}
   \affiliation{Collaborative Innovation Centre of Advanced Microstructures, Nanjing University, Nanjing 210093, China}
   
   \author{Xiaofeng Xu}
   \affiliation{Department of Applied Physics, Zhejiang University of Technology, Hangzhou 310023, China}
   
   \author{T. Tamegai}
   \affiliation{Department of Applied Physics, The University of Tokyo, Tokyo 113-8656, Japan}
   
   \author{Mingxiang Xu}
   \email{Corresponding author:mxxu@seu.edu.cn}
   \affiliation{School of Physics, Southeast University, Nanjing 211189, China}
   
   \author{Zhixiang Shi}
   \email{Corresponding author:zxshi@seu.edu.cn}
   \affiliation{School of Physics, Southeast University, Nanjing 211189, China}

\date{\today}

\begin{abstract}

Intrinsic pinning is caused by natural pinning centers that occur because of the modulation of the order parameter or weak superconducting layers. Early work has shown that intrinsic pinning generates a high pinning force and critical current density in some layered oxide superconductors. Studying the intrinsic pinning of superconductors is crucial for both fundamental studies and potential applications. Herein, we use torque magnetometry to study angle-resolved in-plane and out-of-plane magnetic torque for a series of high-quality FeSe$_1$$_-$$_x$S$_x$ single crystals. A fourfold torque signal was observed when the magnetic field was within the \textit{ab} plane. We interpret that this fourfold in-plane irreversible torque is from the intrinsic pinning due to combined effects of gap nodes/minimum and twin domains. Additionally, we attributed the observed out-of-plane torque peaks to intrinsic pinning due to the layered structure.

\end{abstract}

\pacs{}
\keywords{}
\maketitle

\subsection*{1. Introduction}

Understanding the mechanism of intrinsic pinning in high \textit{T}$_{\rm{c}}$ layered cuprates \cite{tachiki1989strong, tachiki1989anisotropy, roas1990anisotropy, ishida1997two, awaji2010anisotropy, civale2005identification, dam1999origin, janossy1990contributions} and iron-based \cite{iida2013intrinsic, amigo2017intrinsic, iida2013oxypnictide, moll2013transition, sakoda2018recent, tarantini2016intrinsic} superconductors is essential for fundamental studies and potential applications. Type-\uppercase\expandafter{\romannumeral2} superconductors are characterised by the appearance of thin filaments of normally conducting material in the superconducting (SC) state. Each unit carries a quantised amount of magnetic flux, and is circled by a super current vortex. Usually, the model of the vortex pinning with a ``pinning potential'' could understand the motion of individual vortices, which could drive these vortices
%toward the potential wells and
find the pinning sites \cite{chapman1997vortex}. In practice, these vortex pinning sites are generated by the extrinsic defects (such as impurities \cite{liu2021thermal}, dislocations \cite{dam1999origin}, and twin boundaries (TBs) \cite{roas1990anisotropy}) or intrinsic pinning centers, which could reduce the energy penalty associated with the vortex core. Among them, the extrinsic defects usually generates a disordered potential with random distribution or only limit in certain region, while the intrinsic pinning generates a periodic vortex pinning potential \cite{civale2005identification}. The first quantitative intrinsic pinning analysis was conducted in the layered oxide superconductor YBa$_2$Cu$_3$O$_7$ by Tachiki and Takahashi \cite{tachiki1989strong, tachiki1989anisotropy}. Subsequently, intrinsic pinning was also confirmed in other superconductors \cite{iida2013intrinsic, amigo2017intrinsic, iida2013oxypnictide, moll2013transition, xiao2008pairing, xiao2006angular}. Usually, the SC order parameter or layered structure acts as a natural pinning center, driving a system to easy-to-access intrinsic pinning \cite{tachiki1989strong, tachiki1989anisotropy}, corresponding to a high pinning force, critical current density, and magnetic torque signal \cite{ishida1997two}. Therefore, intrinsic pinning can provide information about the SC gap structure and anisotropy of superconductivity. However, it has been studied only in a few superconductors because it is usually mixed or covered by strong extrinsic pinning.

%Additionally, some extrinsic defects (such as impurities \cite{liu2021thermal}, dislocations \cite{dam1999origin}, and twin boundaries (TBs) \cite{roas1990anisotropy}) also act as pinning centers \cite{janossy1990contributions}. The difference between the two is that intrinsic pinning represents a periodic pinning potential, while extrinsic pinning generates a disordered potential with the random distribution or is only limited in certain region of a sample \cite{civale2005identification}.

\begin{figure*}
\includegraphics[width=43pc]{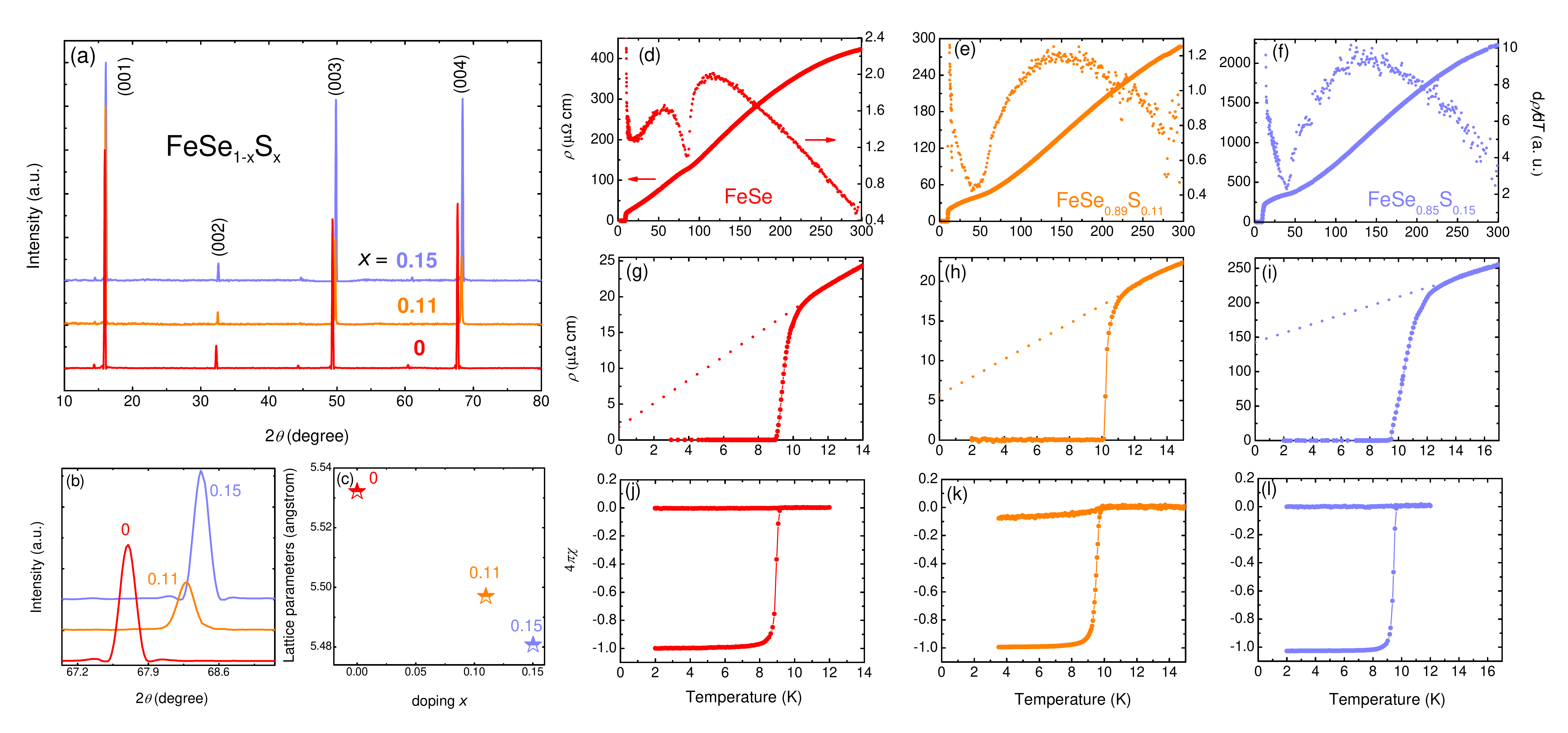}
\begin{center}
\caption{\label{1} (a) Room temperature XRD patterns for the series of FeSe$_1$$_-$$_x$S$_x$ single crystals studied in this paper. (b) Enlarged (004) peaks for all crystals near 68$^{\circ}$. (c) Evolution of the \textit{c}-axis lattice constant as a function of actual doping content \textit{x}. (d)--(f) Temperature dependence of the zero-field resistivity and its first derivative for FeSe$_1$$_-$$_x$S$_x$ single crystals. (g)-(i) show the enlarged plot of the zero-field resistivity data around the SC transition. The dash lines correspond to the power-law fitting $\rho$(T)=$\rho$$_{0}$+\textit{A}\textit{T}$^{\alpha}$ ($\rho$$_{0}$, \textit{A}, and $\alpha$ are the fitting parameters). (j)-(l) Temperature dependence of the zero-field cooling and field cooling magnetic susceptibility $\chi$ after considering the demagnetization effect.}
\end{center}
\end{figure*}

Probing the intrinsic pinning requires a bulk technique with angular resolution. Magnetic torque (\textbf{\textit{$\tau$}} = \textbf{\textit{M}} $\times$ \textbf{\textit{H}}) is a powerful thermodynamic tool used to obtain bulk information and is sensitive in probing any anisotropic susceptibility, which is defined as the first derivative of the free energy with respect to angle (\textbf{\textit{$\tau$}} = - \textit{$\partial$}\textit{F} / \textit{$\partial$}\textit{$\theta$}). This method could provide a direct way to study interesting issues, such as the symmetry of the SC order parameter \cite{xiao2008pairing, willemin1998pairing, ishida1997two} and the anisotropy of intrinsic pinning \cite{ishida1997two, willemin1998strong, willemin1998pairing, xiao2006angular, janossy1990contributions}, \textit{et al}..

The target material in this study is the prototypical 11-family iron chalcogenide FeSe$_1$$_-$$_x$S$_x$ \cite{sun2016electron, hosoi2016nematic, watson2015suppression, abdel2015superconducting}. In \textit{parent} FeSe, it has the simplest PbO-type structure, composed of only Fe-Se layers, and shows superconductivity without further doping \cite{hsu2008superconductivity}. The SC gap is strongly anisotropic \cite{song2011direct, kasahara2014field, watashige2015evidence, bourgeois2016thermal, watashige2017quasiparticle, sato2018abrupt, hanaguri2018two, sun2017gap} with nodes or deep minima \cite{xu2016highly, sprau2017discovery, hashimoto2018superconducting, liu2018orbital, sun2017gap, sun2017symmetry}.
%It has a very simple crystal structure with only stacked FeSe layers \cite{hsu2008superconductivity}.
Meanwhile, it has also been reported to be a very clean system \cite{zhou2021disorder, kasahara2014field, kasahara2020evidence, sun2015critical}. Therefore, FeSe could be an ideal platform for the investigation of the intrinsic pinning \cite{putilov2019vortex}. Recently, some studies construct the phase diagram of FeSe$_{1-x}$S$_{x}$ by using a hydrothermal ion release/introduction technique \cite{yi2021hydrothermal} and the Chemical Vapor Transport (CVT) method \cite{hosoi2016nematic}, respectively. In comparison, these FeSe$_{1-x}$S$_{x}$ (0 $\leq$ \textit{x} $\leq$ 0.2) crystals obtained from the CVT method show higher quality, which are more suitable for studying their intrinsic pinning properties.

%Upon isovalent S substitution at Se sites, the structural transition \textit{T}$_{\rm{s}}$ is gradually suppressed and \textit{T}$_{\rm{c}}$ displays a domelike behavior \cite{hosoi2016nematic}.

Here, we systematically examined the angle-resolved magnetic torque of a series of FeSe$_1$$_-$$_x$S$_x$ (\textit{x} = 0, 0.11, 0.15) single crystals. Fourfold in-plane and twofold out-of-plane magnetic torque peaks were observed at particular magnetic field directions. By analyzing the obtained results, we considered that a combination of gap nodes/minimum and twin domains create the intrinsic pinning responsible for the in-plane magnetic torque peaks. Concurrently, the layered structure are responsible for the out-of-plane magnetic torque peaks.

\subsection*{2. Experiment details}

The single crystals of FeSe$_1$$_-$$_x$S$_x$ (\textit{x} = 0, 0.11, 0.15) studied here were synthesized using the vapor transport method \cite{sun2016electron}. The actual chemical composition was determined via energy dispersive x-ray spectroscopy (EDX). The crystallographic structure was characterized by x-ray diffraction (XRD) at room temperature using a Rigaku diffractometer with Cu \textit{K$\alpha$} radiation. The direction of the crystal axes were determined by single-crystal XRD. Zero-field cooling (ZFC) and field cooling (FC) dc magnetization modes were employed to probe the SC transition and volume fraction using a MPMS-7 system. Electronic transport measurements were performed in a physical property measurement system (PPMS, Quantum Design). High-field transport measurements were carried out in water-cooled magnet with the steady fields up to 38 T at the High Magnetic Field Laboratory of Chinese Academy of Sciences by using standard a.c. lock-in techniques. The torque magnetometry was performed using a piezoresistive technique to measure the torsion of the torque lever in the PPMS system \cite{beck1992torque, gray1990giant}. In this process, if the torsion or twisting of the torque lever is too large, it may move the sample. In some serious cases, the sample even falls down from the stage. Therefore, we must check if the crystal still in the same position after every measurement to ensure that the data were obtained correctly \cite{xu2014electronic}.

%After measurement, the crystal was checked still in the same position without any movement \cite{xu2014electronic}.

%And the single crystal was mounted onto the cantilever using grease. During the torque measurements, we found that the crystal is very easy to be influenced by the force of magnetic torque when the field is rotated in the \textit{ab} plane. Therefore, we add the grease and check the sample not to have moved after every measurement \cite{xu2014electronic}.

\begin{figure*}
\includegraphics[width=40pc]{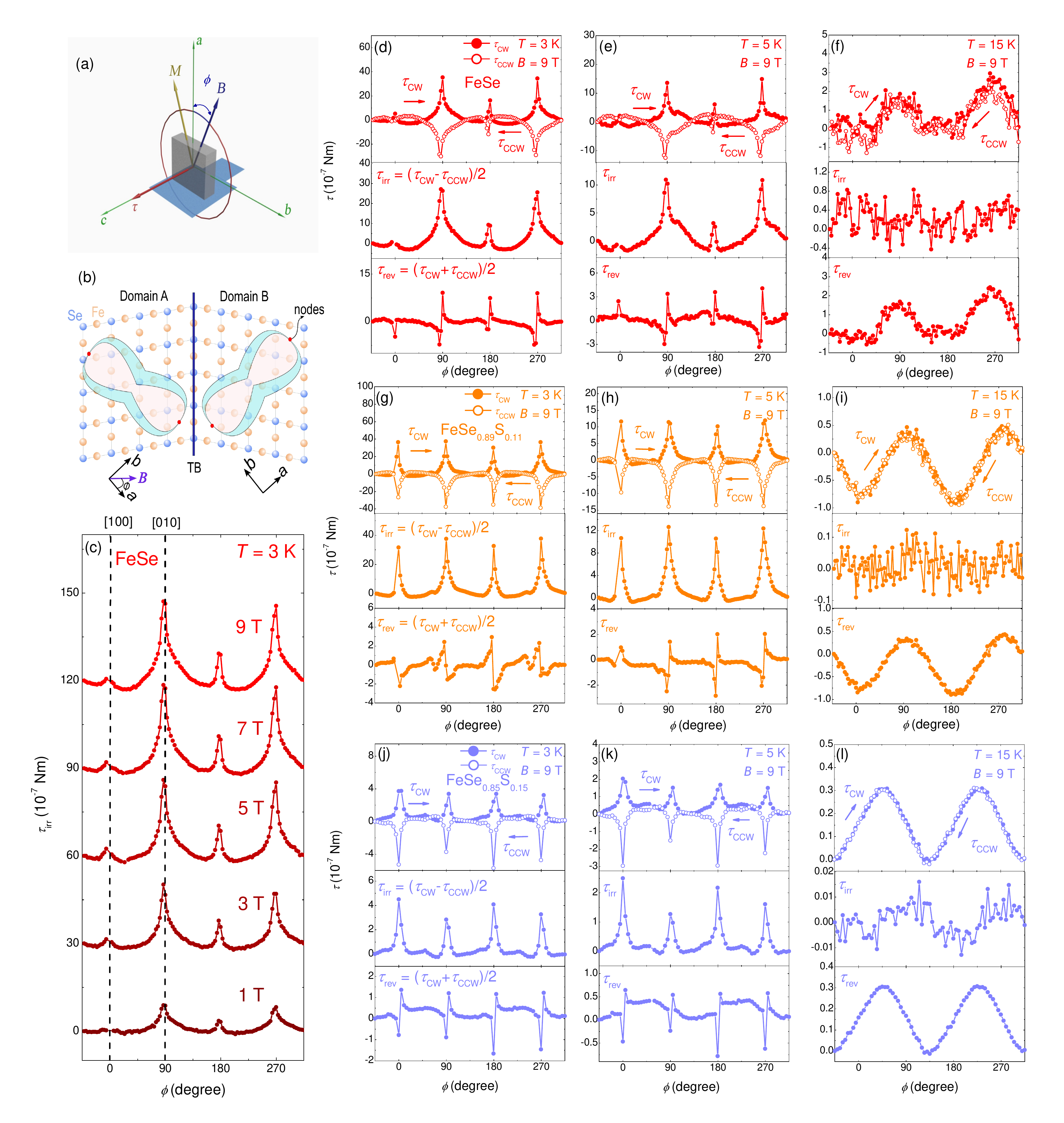}
\begin{center}
\caption{\label{2} (a) Schematic experimental configuration for in-plane magnetic torque measurements. (b) In-plane schematic view of the atomic arrangement and gap structure in domains A and B sandwiching a TB. (c) Angular dependence of irreversible torque \textit{$\tau$}$_{irr}$ under various fields at \textit{T} = 3 K when the field is rotated in the \textit{ab} plane of FeSe. The angle-dependent in-plane clockwise torque \textit{$\tau$}$_{\rm{CW}}$, counterclockwise torque \textit{$\tau$}$_{\rm{CCW}}$, and irreversible torque \textit{$\tau$}$_{irr}$ of (d)-(f) FeSe, (g)-(i) FeSe$_{0.89}$S$_{0.11}$, and (j)-(l) FeSe$_{0.85}$S$_{0.15}$.}
\end{center}
\end{figure*}

\subsection*{3. Results and discussion}

The XRD patterns and their log-scale plot of a series of FeSe$_1$$_-$$_x$S$_x$ single crystals are shown in Fig. 1(a) and Fig. S4 \cite{supplement}, respectively. Only the (00\textit{$\ell$}) peaks are observed, which can be well indexed based on a tetragonal structure with the \textit{P}4/\textit{nmm} space group. The positions of the (00\textit{$\ell$}) peaks were found to shift systematically to higher angles with increasing the S content, which can be seen more clearly in the enlarged plot of the (004) peaks shown in Fig. 1(b). Meanwhile, the extracted \textit{c}-axis lattice parameters are close to the previous studies \cite{moore2015evolution, chareev2018single, yi2021hydrothermal, abdel2015superconducting}, as shown in Fig. 1(c).

\begin{figure*}
\includegraphics[width=40pc]{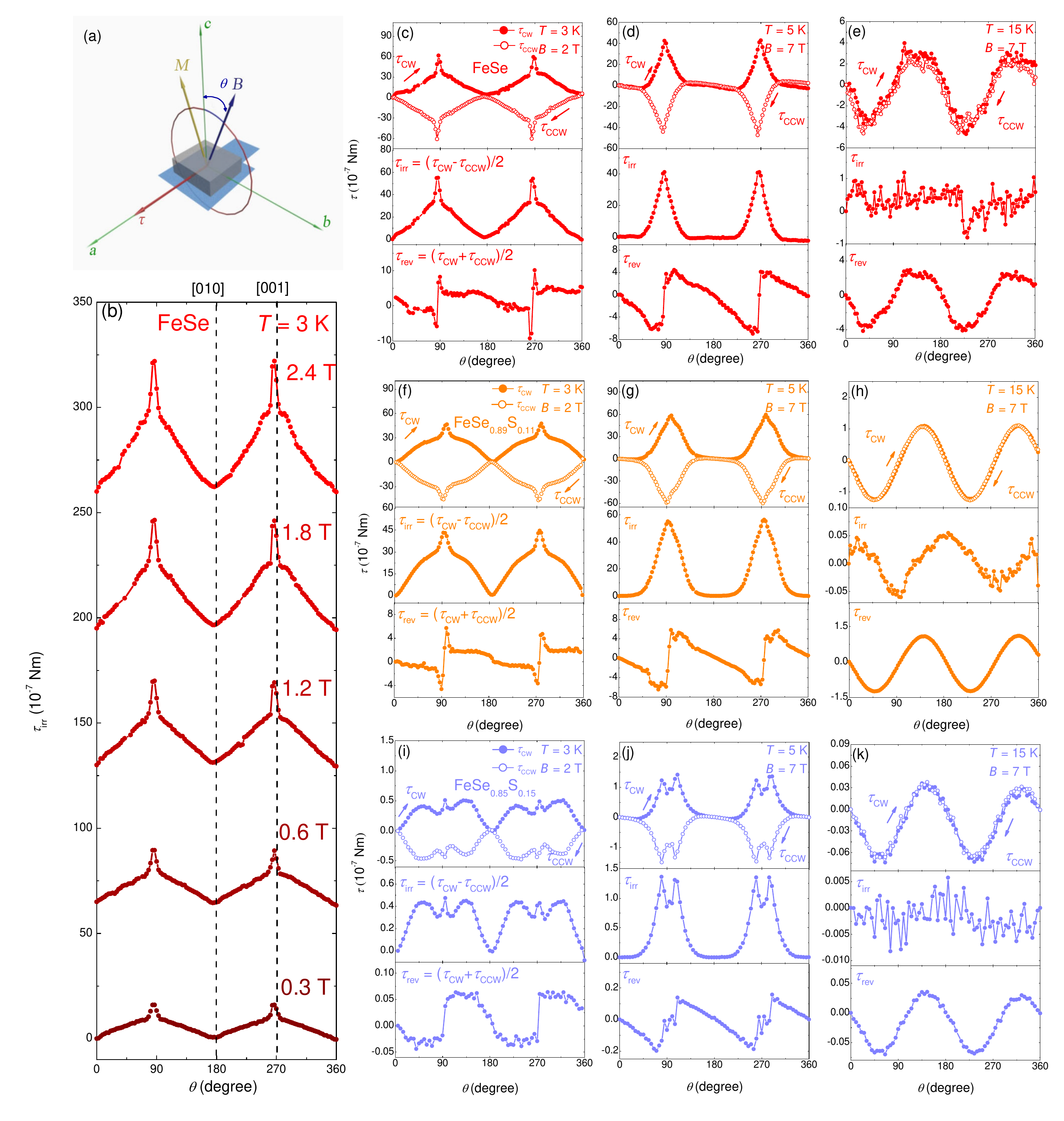}
\begin{center}
\caption{\label{3} (a) Schematic experimental configuration for out-of-plane magnetic torque measurements. (b) Angular dependence of irreversible torque \textit{$\tau$}$_{irr}$ under various fields at \textit{T} = 3 K when the field is rotated out of the \textit{ab} plane of FeSe. The angle-dependent out-of-plane clockwise torque \textit{$\tau$}$_{\rm{CW}}$, counterclockwise torque \textit{$\tau$}$_{\rm{CCW}}$, and irreversible torque \textit{$\tau$}$_{irr}$ of (c)-(e) FeSe, (f)-(h) FeSe$_{0.89}$S$_{0.11}$, and (i)-(k) FeSe$_{0.85}$S$_{0.15}$.}
\end{center}
\end{figure*}

%These results indicate that the S atoms are successfully incorporated into the system.

Figures 1(d)-(f) depict the zero-field temperature-dependent resistivity for FeSe$_1$$_-$$_x$S$_x$ single crystals. The residual resistivity $\rho$$_{0}$ is determined by using the power-law fitting $\rho$(T)=$\rho$$_{0}$+\textit{A}\textit{T}$^{\alpha}$ ($\rho$$_{0}$, \textit{A}, and $\alpha$ are the fitting parameters) from normal state data to zero temperature as shown by the dashed lines in Figs. 1(g)--1(i). The obtained $\rho$$_{0}$ are $\sim$ 1.83 $\mu$$\Omega$ cm (FeSe), 5.69 $\mu$$\Omega$ cm (FeSe$_{0.89}$S$_{0.11}$), and 142.85 $\mu$$\Omega$ cm (FeSe$_{0.85}$S$_{0.15}$), respectively. The residual resistivity ratio (RRR), defined as $\rho$$_{300 K}$/$\rho$$_{0}$, is estimated to be $\sim$ 234 for FeSe, $\sim$ 51 for FeSe$_{0.89}$S$_{0.11}$, and $\sim$ 16 for FeSe$_{0.85}$S$_{0.15}$, respectively. Meanwhile, the value of RRR has also been calculated by using the expression of $\rho$$_{300 K}$/$\rho$$_{T\rm{c}}$, which is $\sim$ 25 for FeSe, $\sim$ 16 for FeSe$_{0.89}$S$_{0.11}$, and $\sim$ 10 for FeSe$_{0.85}$S$_{0.15}$. It is reasonable that the value of RRR decreases when introducing more sulphur in Se site.
%Furthermore, we also try to compare these results with the recent studies \cite{zhou2021disorder, sun2017gap, sun2018disorder}, which indicate that the crystals used in current study maintain a relatively higher quality.}
The structural transition \textit{T}$_{\rm{s}}$ could be seen more clearly by plotting the first-order derivative of the resistivity (d\textit{$\rho$}/d\textit{T}). Upon isovalent S substitution, a domelike dependence of \textit{T}$_{\rm{c}}$ on S doping has been observed \cite{hosoi2016nematic, yi2021hydrothermal}, which corresponds to a maximum SC transition temperature \textit{T}$_{\rm{c}}$ of $\sim$ 10.1 K at the doping level of \textit{x} $\sim$ 0.11 (Figs. 1(g)-(i)). The large diamagnetic signal and sharp SC transition were confirmed in all crystals (see Figs. 1(j)-(l)). \textit{T}$_{\rm{c}}$ for each sample is defined as the onset of the separation between FC and ZFC curves and it agrees with the onset of resistivity in Figs. 1(g)-(i).

Angular dependence of the in-plane torque \textit{$\tau$}(\textit{$\phi$}) and out-of-plane torque \textit{$\tau$}(\textit{$\theta$}) were investigated for series of FeSe$_1$$_-$$_x$S$_x$ single crystals with angles schematically shown in Fig. 2(a) and Fig. 3(a), respectively. Both the in-plane and out-of-plane magnetic torque were measured clockwise \textit{$\tau$}$_{\rm{CW}}$ (solid circles) and counterclockwise \textit{$\tau$}$_{\rm{CCW}}$ (open circles). The reversible torque \textit{$\tau$}$_{rev}$ was estimated as (\textit{$\tau$}$_{\rm{CW}}$ + \textit{$\tau$}$_{\rm{CCW}}$)/2, while the irreversible torque \textit{$\tau$}$_{irr}$ was given by (\textit{$\tau$}$_{\rm{CW}}$ - \textit{$\tau$}$_{\rm{CCW}}$)/2 \cite{ishida1997two, willemin1998pairing, xiao2008pairing} (see Figs. 2 and 3). When FeSe$_1$$_-$$_x$S$_x$ are in the normal state (\textit{T} = 15 K), the in-plane and out-of-plane torque data exhibit a twofold symmetry and the clockwise torque \textit{$\tau$}$_{\rm{CW}}$ and counterclockwise torque \textit{$\tau$}$_{\rm{CCW}}$ show the same behavior, indicating a fully reversible torque. Correspondingly, the irreversible torque \textit{$\tau$}$_{irr}$ is almost absent. The full reversibility confirms that there is no pinning contribution. As the temperature decreases, when these samples are in the SC state, a significant difference is observed between \textit{$\tau$}$_{\rm{CW}}$ and \textit{$\tau$}$_{\rm{CCW}}$ in both the in-plane and out-of-plane torque data. Meanwhile, the irreversible torque \textit{$\tau$}$_{irr}$ is enhanced and becomes dominant at low temperatures.

%Unlike the behavior of the out-of-plane magnetic torque, t
The in-plane irreversible torque of the three crystals display sharp peaks with maxima at 0$^{\circ}$, 90$^{\circ}$, 180$^{\circ}$, and 270$^{\circ}$ in the SC mixed state (see Fig. 2). The evolution of the fourfold magnetic torque under various fields at 3 K for FeSe is shown in Fig. 2(c).
%Similar to the out-of-plane torque,
The peaks become strong with increasing the field. We also notice that the peaks of the irreversible torque occurs when the field is applied along the \textit{a} axis or the \textit{b} axis (Fig. 2(c)). Similar results have also been observed in CeCoIn$_{5}$ \cite{xiao2008pairing} and untwinned YBa$_{2}$Cu$_{3}$O$_{7}$ \cite{ishida1997two} superconductors. In untwinned YBa$_{2}$Cu$_{3}$O$_{7}$, the fourfold torque peaks have also been observed along \textit{a} or \textit{b} axis, which have been considered that the free-energy minima due to gap nodes work as intrinsic pinning centers, making vortices ensconce themselves in local minima of free energy \cite{ishida1997two}. These results support the \textit{d}$_{\textit{x}^2-\textit{y}^2}$ gap symmetry in YBa$_{2}$Cu$_{3}$O$_{7}$ \cite{ishida1997two}.
%\textcolor{red}{Furthermore, in experiments, we also found that the position of these peaks are associated only with the crystal axis rather than the crystal shape.}
%Furthermore, some recent studies confirm the SC gap symmetry of CeCoIn$_{5}$ to be \textit{d}$_{\textit{x}^2-\textit{y}^2}$ \cite{allan2013imaging, sakakibara2016angle}. However, interestingly, the torque peaks show a phase shift of $\pi$/4 relative to the YBa$_{2}$Cu$_{3}$O$_{7}$ \cite{xiao2008pairing}, indicating the in-plane irreversible torque along [110] in CeCoIn$_{5}$.

FeSe comprises one hole-type \textit{$\alpha$} band located at the \textit{$\Gamma$} point and two electron-type bands \textit{$\delta$} and \textit{$\varepsilon$} at the \textit{M} point \cite{terashima2014anomalous, watson2016evidence}. Some previous studies provide evidence for the presence of deep minima or nodes in a wide \textit{x} range of FeSe$_1$$_-$$_x$S$_x$ crystals (\textit{x} $\leq$ 0.17) \cite{sato2018abrupt, hanaguri2018two}. Here our results confirm that these in-plane torque peaks are also observed in the three crystals (0 $\leq$ \textit{x} $\leq$ 0.15). Therefore, it is reasonable to assume that a possible scenario with gap nodes, which is similar to the situations in untwinned YBa$_{2}$Cu$_{3}$O$_{7}$ \cite{ishida1997two} superconductors. However, Bogoliubov quasiparticle interference measurements found that the gaps from the \textit{$\varepsilon$}($\triangle$$_{\textit{$\varepsilon$}}$) and \textit{$\alpha$}($\triangle$$_{\textit{$\alpha$}}$) bands are twofold symmetric \cite{sprau2017discovery}, which is inconsistent with the behavior of the observed fourfold torque signal, unless we consider the presence of twin domains. By carefully checking our magnetic torque results, we also noticed that the extracted fourfold irreversible torque peaks is not exactly the same. The amplitude of the magnetic torque peaks at 0$^{\circ}$ and 180$^{\circ}$ are significant larger than these peaks at 90$^{\circ}$ and 270$^{\circ}$, which seems to correspond to two different sets of peaks with twofold symmetry (see Fig. 2). Combining these observations with our previous report \cite{sun2017gap}, we realized that the fourfold torque should be related to the presence of twin domains.

In order to identify the twin domains, we check the topology surface for FeSe single crystal by using the polarized light at temperatures above and below the structure transition temperature. The stripe-like structures are observed only at temperature ($\sim$ 5 K) below \textit{T}$_{\rm{s}}$, as marked by the red solid lines in Fig. S1 \cite{supplement}. We also noticed that the neighboring domains A and B are rotated by 90$^{\circ}$ (see Fig. S1 (b) \cite{supplement}). Therefore, the corresponding gap functions are also supposed to shift by 90$^{\circ}$ between the two domains (see Fig. 2(b)). Then, considering the superposition of the two sets of gap functions, the twofold signal will turn into fourfold just as we observed. The asymmetric peak amplitude has been observed in all FeSe$_1$$_-$$_x$S$_x$ crystals, which could be considered to originate from the fact that the ratio of the two kinds of domains differs from 1:1 based on the above discussions. For these FeSe-based materials, this kind of the fourfold symmetry has been reported in angular dependent heat capacity studies \cite{sun2017gap}, which were also attributed to the effect of the twin domains. In order to rule out the influence of twin domains, future efforts on the untwined or single domained FeSe crystal are hopefully to solve this issue.

The low-temperature out-of-plane torque results are presented in detail in Fig. 3 for the three crystals. Some sharp irreversible torque peaks are observed close to 90$^\circ$ and 270$^\circ$ at 3 K when the magnetic field crosses the FeSe planes (\textit{B} $\parallel$ \textit{ab} plane) in the SC mixed state for all the three crystals (see Fig. 3). Such sharp peaks are suppressed with increasing temperature up to 5 K, then a broad peak (ranging from 45$^\circ$ to 135$^\circ$) is observed (see the middle panel in Figs. 3(d), 3(g), and 3(j)). For these sharp peaks, we discuss the possible scenario of intrinsic pinning between two adjacent FeSe layers. In general, certain two-dimensional materials can easily form an alternate stacking of strong and weak SC layers, e.g. YBa$_{2}$Cu$_{3}$O$_{7}$ \cite{tachiki1989strong, tachiki1989anisotropy}. When we inject vortices parallel to these SC layers, in order to keep the loss of the SC condensation energy minimum, the weak SC layers usually work as natural pinning centers to be preferentially accommodated by vortices, generating a high pinning force. In the case of FeSe, the coherence length $\xi$$_{\rm{c}}$ is estimated as $\sim$ 8.7 ${\rm{\AA}}$ (More details of the estimation of $\xi$$_{\rm{c}}$ are shown in Figs. S2 and S3 \cite{supplement}), which is still larger but comparable to the lattice constant \textit{c} ($\sim$ 5.52 ${\rm{\AA}}$). Here, we want to point out that the value of $\xi$$_{\rm{c}}$ is actually over-estimated because the multi-band effect is not considered. In other word, the actual value of $\xi$$_{\rm{c}}$ for FeSe should be closer to the value of \textit{c}. In this case, the superconductivity in the inter-layer part will be weaker than that in the FeSe layer, which acts as intrinsic pinning centers. Compared to the case of YBa$_{2}$Cu$_{3}$O$_{7}$, such intrinsic pinning is weaker. It can be observed only at low temperatures due to shorter SC coherence length. In a similar compound FeTe$_{0.5}$Se$_{0.5}$, the intrinsic pinning from the inter-layer structure has been observed by the measurements of the angle-dependence of \textit{J}$_{\rm{c}}$ \cite{iida2013intrinsic}. Base on the above discussions, future efforts on the accurate estimation of the SC coherence length are important for understanding the issue.

On the other hand, the broad peak ranging from 45$^\circ$ to 135$^\circ$ may be from the surface pinning, because the specific surface area of the crystals used for the torque measurements is relatively large (typical size 1 $\times$ 0.5 $\times$ 0.06 mm$^{3}$). Besides, a shoulder-like behavior is observed only in the S-doped crystals, and can be witnessed more clearly in the crystal of \textit{x} = 0.15 (see the middle panel in Figs. 3(i) and 3(j)). The position and shape of the shoulder-like structure change with temperatures and fields, which is obviously different from those of the intrinsic pinning and surface pinning. Such behavior is usually attributed to the coexistence and competition of several extrinsic pinning centers. In FeSe$_1$$_-$$_x$S$_x$, the extrinsic pinning due to TBs \cite{sun2016electron, sun2017gap, watashige2015evidence} and point-like defects \cite{sun2015critical, sun2015enhancement} have been confirmed. The extrinsic pinning can be only clearly seen in the FeSe$_{0.85}$S$_{0.15}$, which indicates that the extrinsic pinning effect could be enhanced when S are introduced on Se sites, and they represent a competing nature with intrinsic pinning.
%Furthermore, we can simulate the torque data from the magnetic hysteresis loops (MHLs) based on the formula of $\tau$ = \textit{B}sin($\theta$)\textit{M}$_{c}$[\textit{B}cos($\theta$)] \cite{fruchter1989torque}. The broad hump-like structures observed in FeSe$_{0.85}$S$_{0.15}$ should lead to the second-peak effect in MHLs. However, the second-peak has not been observed in our previous work \cite{sun2016effect}, which may be due to the different sensitivity between MHLs and magnetic torque measurements, or the sample-dependent disorders.

Furthermore, in addition to the intrinsic pinning effect in the superconducting state, the electronic nematicity is also an important and interesting issue in FeSe-based superconductor. Unlike other iron-based families \cite{xu2014electronic, kasahara2012electronic}, FeSe exhibits an electronic nematic order without accompanying antiferromagnetic order \cite{hosoi2016nematic}. Some previous studies have confirmed that the rotational symmetry breaking of the electronic nematicity could be revealed by the in-plane anisotropy susceptibility \cite{xu2014electronic, kasahara2012electronic}. Therefore, it is worth trying to detect the nematic phase in FeSe$_{1-x}$S$_{x}$ crystals by using the magnetic torque technology. Meanwhile, some results also show the anisotropy torque signal of the normal state may include two distinct sources, one from external impurities and the other from the nematic phase \cite{xu2014electronic}. Therefore, some experimental details and analysis method need to be repeatedly confirmed in future studies, ensuring that the intrinsic nematic signal is obtained.

\subsection*{4. Conclusions}
%\section{CONCLUSIONS}
In summary, we observed fourfold and twofold torque peaks at particular magnetic field directions in the SC state for FeSe$_1$$_-$$_x$S$_x$ single crystals.  We interpret the in-plane fourfold irreversible torque to the intrinsic pinning, which is generated by a combined effects of gap nodes/minima and twin domains. We attribute out-of-plane twofold torque peaks to intrinsic pinning due to the layered structure.

\subsection*{Credit author statement}

Nan Zhou: Conceptualization, Data Curation, and Writing-Original draft preparation. Yue Sun, T. Tamegai, T. Sakakibara and Xiaofeng Xu: Methodology and Writing-Review \& Editing. Q. Hou, X. Z. Xing, C. Q. Xu, C. Y. Xi, Z. S. Wang, Y. F. Zhang, Y. Q. Pan, and B. Chen: Formal analysis and Resources.
Mingxiang Xu and X. Luo: Supervision, Writing-Review \& Editing. Zhixiang Shi and Yuping Sun: Project administration.

\subsection*{Declaration of competing interest}

The authors declare that they have no known competing financial interests or personal relationships that could have appeared to influence the work reported in this paper.

\subsection*{Data availability}

The authors do not have permission to share data.

\subsection*{Acknowledgements}
%\section*{ACKNOWLEDGMENTS}
A portion of this work was performed on the Steady High Magnetic Field Facilities, Chinese Academy of Sciences, and supported by the High Magnetic Field Laboratory of Anhui Province. This work was partly supported by the National Natural Science Foundation of China (Grant No. 12204487, No. 12204265, No. U1932217, No. 12274369 and No. 11974061), the National Key R$\&$D Program of China (Grant No. 2018YFA0704300 and No. 2021YFA1600201), the Strategic Priority Research Program (B) of the Chinese Academy of Sciences (Grant No. XDB25000000), the Natural Science Foundation of Jiangsu Province of China (Grant No. BK20201285), and JSPS KAKENHI (No. JP20H05164, No. JP19K14661, and No. JP17H01141). Fundamental Research Funds for the Central Universities.

N. Z. and Y. S. contributed equally to this paper.

\bibliographystyle{apsrev4-2}
\bibliography{FeSeStorquepaper}

\pagebreak
\newpage
%\widetext
\onecolumngrid
\begin{center}
	\textbf{\huge Supplemental information}
\end{center}
\vspace{1cm}
\onecolumngrid
%%%%%%%%%% Merge with supplemental materials %%%%%%%%%%
%%%%%%%%%% Prefix a "S" to all equations, figures, tables and reset the counter %%%%%%%%%%
\setcounter{equation}{0}
\setcounter{figure}{0}
\setcounter{table}{0}

\makeatletter
\renewcommand{\theequation}{S\arabic{equation}}
\renewcommand{\thefigure}{S\arabic{figure}}

\begin{figure*}[htp]
\includegraphics[width=36pc]{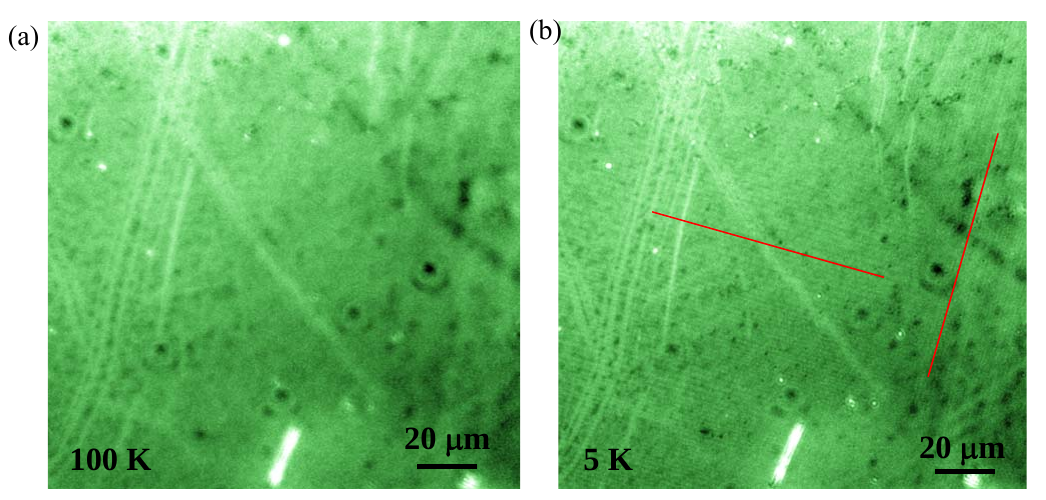}
\caption{\label{S2} The optical images of the surface of a FeSe single crystal at the same position observed in 50 $\times$ microscopy at (a) 100 K and (b) 5 K, respectively.}
\end{figure*}

\begin{figure*}
\includegraphics[width=30.5pc]{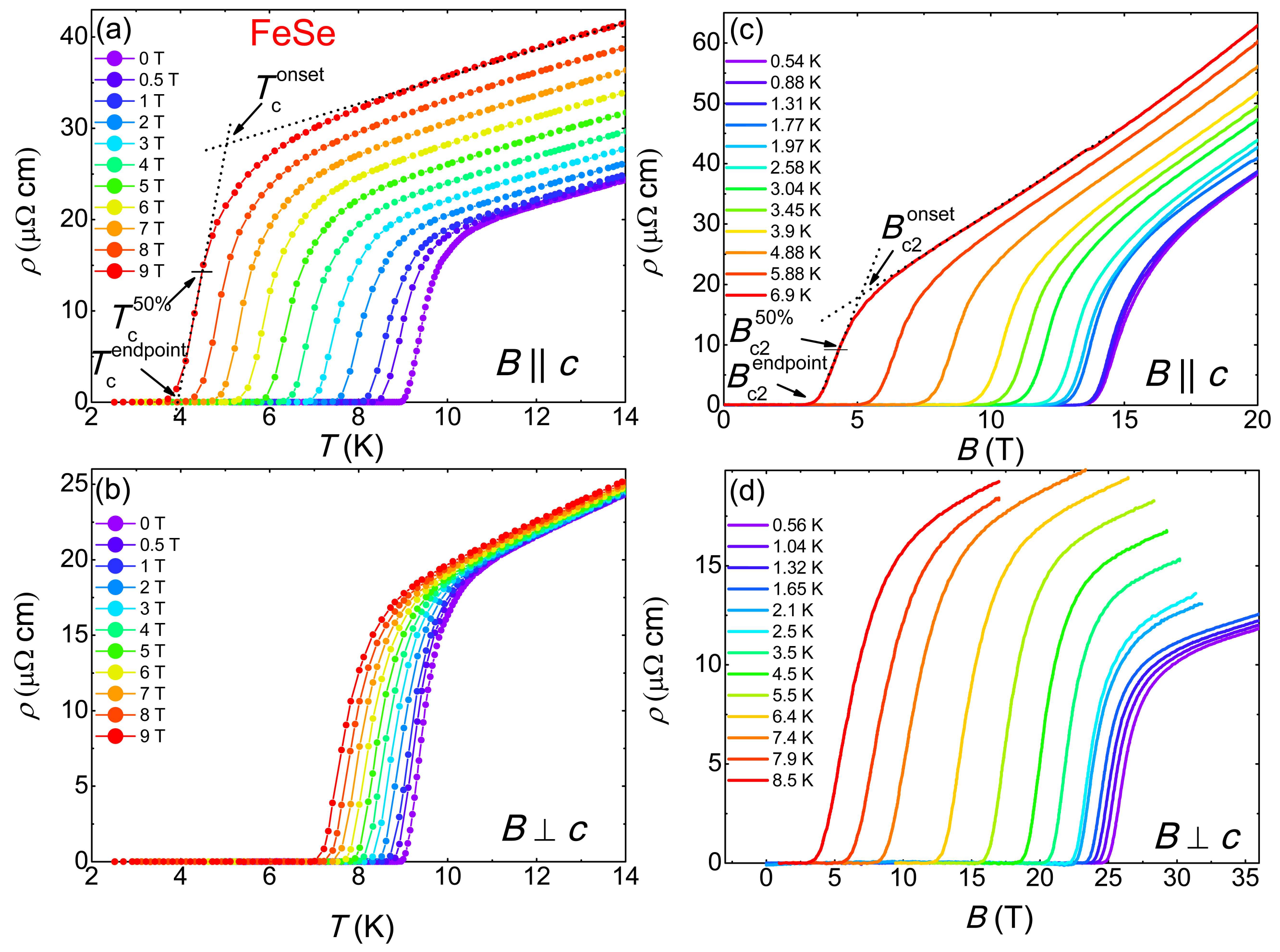}
\caption{\label{S3} Temperature dependence of the resistivity for FeSe single crystal with (a) \textit{B} $\parallel$ \textit{c} and (b) \textit{B} $\perp$ \textit{c} under fields from 0 to 9 T. Magnetic field dependence of the resistivity for FeSe single crystal with (c) \textit{B} $\parallel$ \textit{c} and (d) \textit{B} $\perp$ \textit{c} under fields from 0 to 38 T.}
\end{figure*}

\begin{figure*}
\includegraphics[width=20pc]{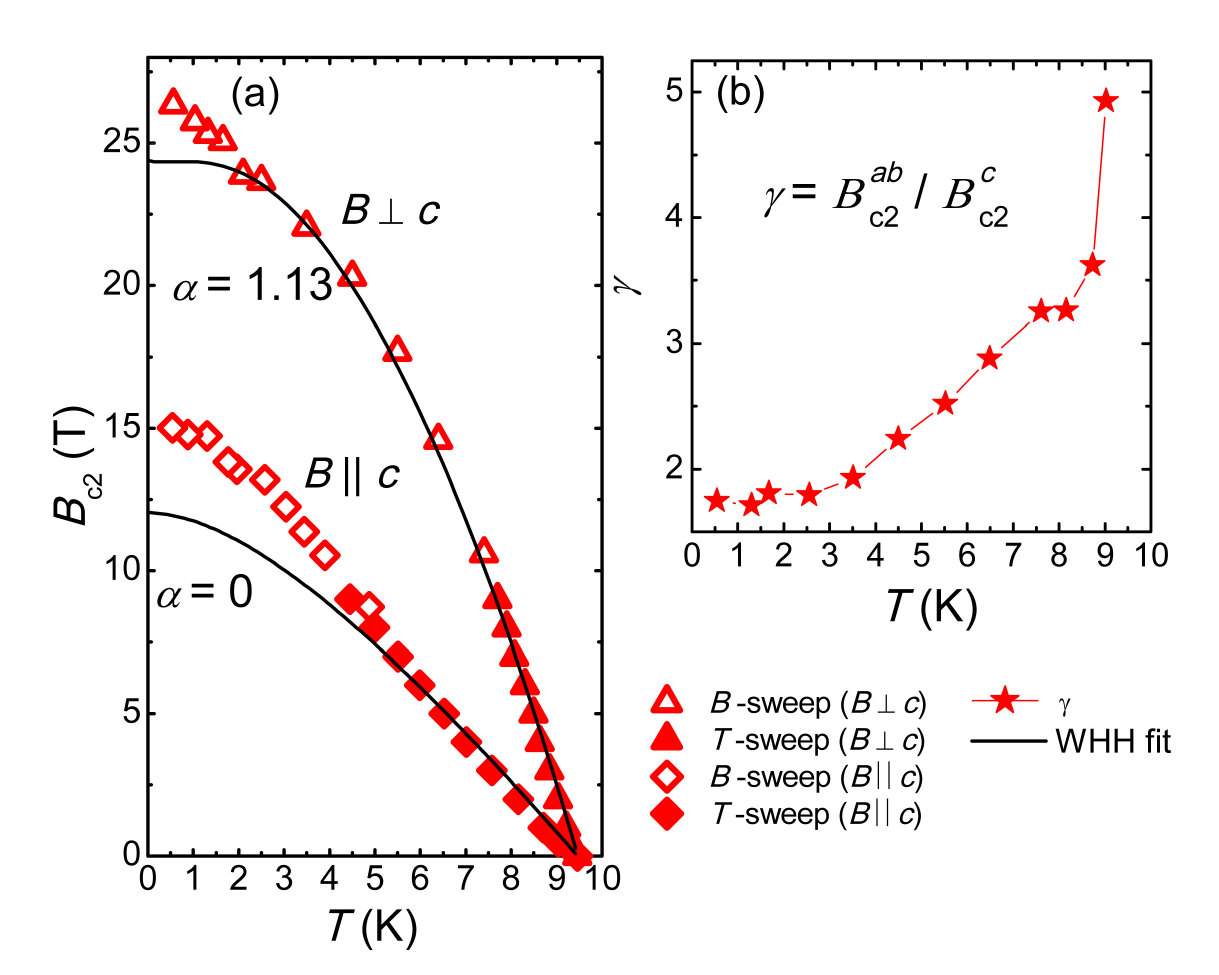}
\begin{center}
\caption{\label{S4} (a) Temperature dependent upper critical field \textit{B}$_{\rm{c}2}$(\textit{T}) for FeSe single crystal. Symbols of the diamond and circle represent the case of \textit{B} $\parallel$ \textit{c} and \textit{B} $\parallel$ \textit{ab} plane, respectively. The closed and open symbols correspond to the \textit{T}- and \textit{B}-sweep resistivity measurements, respectively. The determination of \textit{B}$_{\rm{c}2}$ by the criteria of 50\% of the SC transition is adopted for discussion. Some recent results note that the behavior of the upper critical field in FeSe could be greatly affected by some extra factors under high field region. Meanwhile, we also found that the experimental data significantly deviates from the Werthamer-Helfand-Hohenberg (WHH) curves (black lines) both in \textit{B} $\parallel$ \textit{c} and \textit{B} $\parallel$ \textit{ab} plane. Therefore, in order to eliminate these high field effects, we estimate the \textit{$\xi$}$_{\textit{c}}$ by only considering the parameters near \textit{T}$_{\rm{c}}$. We obtain \textit{$\xi$}$^{\textit{ab}}_{0}$ $\sim$ 43.7 ${\rm{\AA}}$ by using the expressions \textit{$\xi$}$^{\textit{ab}}_{0}$ = ($\Phi$$_0$/2\textit{$\pi$}\textit{T}$_{\rm{c}}$\textit{B}$^{'}$)$^{1/2}$ ($\Phi$$_0$ = 2.07$\times$10$^{-15}$ Wb is the magnetic flux quantum, \textit{T}$_{\rm{c}}$ $\sim$ 9.46 K, \textit{B}$^{'}$ = \textit{d}\textit{B}$^{\textit{c}}_{\rm{c}2}$/\textit{d}\textit{T} $\sim$ 1.83 TK$^{-1}$). Subsequently, the coherence length \textit{$\xi$}$_{\textit{c}}$ $\sim$ 8.7 ${\rm{\AA}}$ is determined by using the expressions \textit{$\xi$}$_{\textit{c}}$ = \textit{$\xi$}$^{\textit{ab}}_{0}$\textit{B}$^{\textit{c}}_{\rm{c}2}$/\textit{B}$^{\textit{ab}}_{\rm{c}2}$ (\textit{B}$^{\textit{c}}_{\rm{c}2}$/\textit{B}$^{\textit{ab}}_{\rm{c}2}$ $\sim$ 1/5 near \textit{T}$_{\rm{c}}$), which is comparable to the lattice constant \textit{c} ($\sim$ 5.52 ${\rm{\AA}}$). (b) Temperature dependence of the anisotropy parameter \textit{$\gamma$} = \textit{B}$^{\textit{ab}}_{\rm{c}2}$/\textit{B}$^{\textit{c}}_{\rm{c}2}$.}
\end{center}
\end{figure*}

\begin{figure*}
\includegraphics[width=43pc]{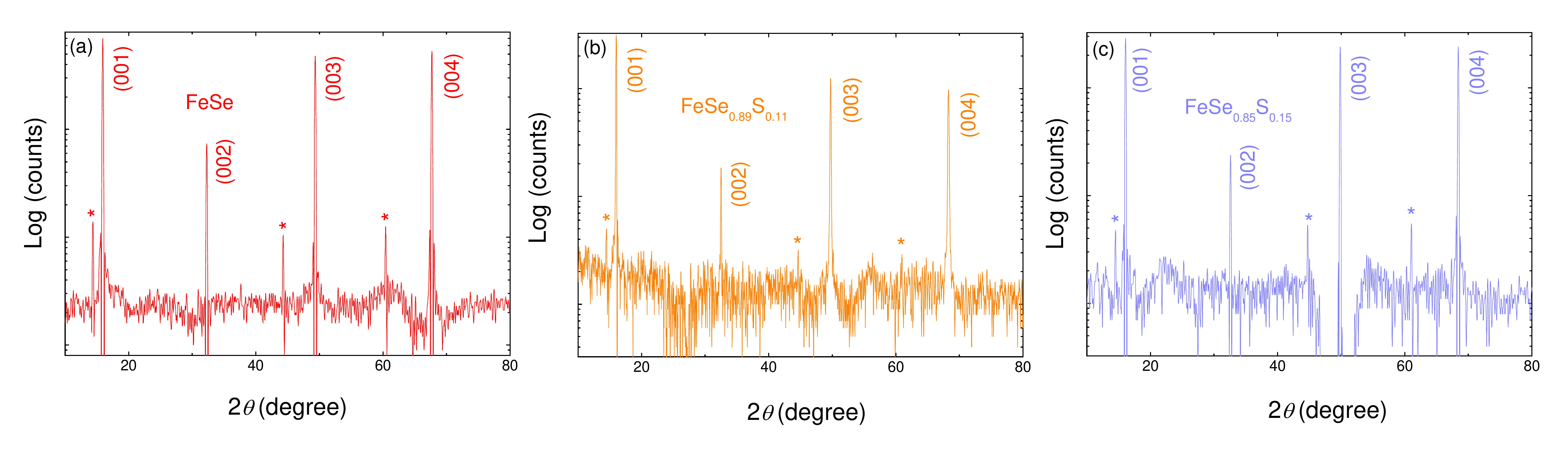}
\caption{\label{S2} The log-scale plot of single-crystal x-ray diffraction pattern for (a) FeSe, (b) FeSe$_{0.89}$S$_{0.11}$, and (c) FeSe$_{0.85}$S$_{0.15}$, respectively. The asterisk corresponds to the \textit{k}$_{\beta}$ peaks.}
\end{figure*}

\end{document}